\documentclass{pinchcr}   

\usepackage{graphicx}       
\usepackage{cite}            
\usepackage{amssymb}
\usepackage[bottom]{footmisc}

\begin{document}


\title{Heterogeneities in amorphous systems under shear}

\author{Jean-Louis Barrat}
\affiliation{Universit\'e Joseph Fourier Grenoble 1;   Laboratoire interdisciplinaire de physique; CNRS,
UMR 5588, 38420  Saint Martin d'H\'eres Cedex,
France \texttt{jean-louis.barrat@ujf-grenoble.fr}}

\author{Ana\"el Lema\^itre}
\affiliation{Universit\'e Paris-Est -- Navier -- CNRS UMR 8205 -- ENPC -- LCPC\\
2 all\'ee K\'epler, 77420 Champs-sur-Marne, France\\
\texttt{anael.lemaitre@ifsttar.fr}}

\maketitle

\preface
The last decade has seen major progresses in studies of elementary mechanisms of deformation in amorphous materials. Here, we start with a review of physically-based theories of plasticity, going back to the identification of ``shear-transformations'' as early as the 70's. We show how constructive criticism of the theoretical models permits to formulate questions concerning the role of structural disorder, mechanical noise, and long-ranged elastic interactions. These questions provide the necessary context to understand what has motivated recent numerical studies. We then summarize their results, show why they had to focus on athermal systems, and point out the outstanding questions.

\section{Introduction}
\label{sec:1}

Rheology and plasticity, although they both investigate the flow
of solid materials,  are generally considered as two separate fields of materials science.
Plasticity deals with the deformation of "hard" solids, characterized by large elastic moduli
(typically in the GPa range). Rheology, on the other hand, deals with much softer materials, such as  colloidal pastes,
foams, or other "complex fluids"\shortcite{Larson}  with moduli that can vary from a few Pa to kPa. In view of these differences, the experimental tools
used to investigate the flow of hard and soft materials  differ widely, whether they are mechanical or involve more indirect
microscopic characterizations.

Still, if one temporarily forgets about the differences in the scale of stress levels,
striking similarities appear in the behavior of these different materials, as illustrated schematically in figure \ref{fig:stress-strain}.
The differences in stress scales are indeed easily  understood in terms of the interactions.  The scale for elastic moduli is an energy per unit volume.
In hard materials, typical energies will be in the range 0.1-1eV, and the typical length scales are of order of nanometers, or even smaller. In softer materials,
the energy scale is often comparable to $k_BT$, and length scales of the order of a tenth of a micron. Finally, the case of foams corresponds to a stress scale set by the surface tension $\gamma$ divided by the typical bubble size. It is then not impossible, that common physical properties can be found in
such widely different systems if the proper elementary units are considered and the appropriate rescalings are made.

In terms of "reduced" parameters, 
however, the experimental conditions may correspond to very different ranges for the various systems investigated. For example,
a foam or a two dimensional "bubble raft"
 is essentially always athermal (the thermal fluctuations are irrelevant compared to the energies involved at the scale of individual bubbles)
so that it should be compared to metallic or polymer glasses at low temperatures. On the other hand, a not too dense colloidal system at room temperature, in which thermal fluctuations are significant,
could be compared to systems close to their glass transition temperature. The same type of considerations apply to time scales and their comparison with the applied deformation rates, which have a strong influence on the stress-strain curve, as sketched in figure \ref{fig:stress-strain}. Typically, the
stress peak $\sigma_{max}$ shown in this figure tends to display a  logarithmic increase  with the deformation rate $\dot{\epsilon}$
and, when ageing is observed, with the age, $t_w$ of the system.

In this chapter, we will limit our considerations to noncrystalline, amorphous materials.
In crystalline materials, flow can be described in terms of dislocation motion,
and although the interaction between these extended defects may lead to a macroscopic behavior similar
to that of amorphous materials \shortcite{Miguel2006}, the underlying microscopic physics is different. For amorphous materials
it will be shown that  flow defects, if they exist,  are localized rather than extended.  We will also
exclude from our considerations the case of  real granular materials, which raise several complications such as the importance of gravity and that of
friction. 

Why is a chapter on sheared materials included in a book on dynamical heterogeneities? 
The standard plasticity or rheology approach is based on macroscopic constitutive equations, 
established  using  symmetry arguments~\shortcite{Lubliner2008}.
These equations  relate the stress and strain  (or strain history) in the system, within a
fully homogeneous, continuous medium description. 
However, the notion of 
dynamical heterogeneities naturally emerges 
when
one attempts to reach an understanding of the microscopic mechanisms that underly the macroscopic behavior.
Is it possible to identify microscopic heterogeneities, that would in some respects play the role assigned to dislocations
in the flow of crystalline materials? What governs the dynamical activity of such heterogeneities, what are their interactions
and correlations, and do they organize on larger scales? From the pioneering experiments of Argon \shortcite{ArgonKuo1979}
emerged the notion of "shear transformations", localized (in space and time) rearrangements which govern the plastic activity~\shortcite{Argon1982}. 
Such local yield events have been very clearly identified in experiments on bubble rafts, in colloidal systems~\shortcite{SchallWeitzSpaepen2007},
as well as in various atomistic simulations of low temperature deformation (see e.g. \shortcite{FalkLanger1998,MaloneyLemaitre2006,TanguyLeonforteBarrat2006}). They are
now believed to constitute the elementary constituent of plastic deformation in amorphous solids at low temperature. However,
their cooperative organisation is far from being understood, although a number of models based of this notion of
elementary event have been developed and studied  analytically at the mean field level or numerically. The discussion of these
microscopic, dynamical heterogeneities will be the core of this chapter.

Another, very different aspect that escapes the purely  macroscopic 
description of flow is
 the frequent experimental observation  that strain in solid materials can take place in a very heterogeneous manner at the macroscopic level.
This phenomenon, described as the existence of  "shear bands" or "strain localisation", takes place both in hard and soft materials, under various conditions of deformation. Instead of being evenly distributed and uniform through the system (affine deformation), the deformation is concentrated inside a localized region of space, typically a two dimensional "shear band" with a finite thickness. Within such  shear bands where the entire
macroscopic deformation is concentrated, the local strain  becomes very large in a short time, eventually leading to material failure for hard materials.
In soft materials, which, in contrast to hard materials, can sustain steady flow when strained
in a pure shear (Couette) geometry, shear bands can  become permanent features of the steady state flow and coexist with immobile parts of the same material. In such cases the flowing part, which is also described as a shear band,  occupies a finite fraction of the sample thickness. While shear bands are clearly macroscopic
features of the flow, understanding the mechanism of their formation and stability necessitates the introduction of auxiliary 
state variables and associated length scales, which in turn could be related to the existence of flow heterogeneities at smaller scales. These aspects will be discussed further in section 1.2.7.

We mention at the end of  this introduction that the focus of our paper will be on simple shear deformation. In most materials, it is expected that
the study of such deformations will allow one to unveil the mechanisms of plasticity in more complex situations, as the plastic deformation proceeds
with little or no density increase. This is true in metallic glasses or soft condensed system, where plastic
deformation often  takes place at almost constant volume, but less in network materials with a lower Poisson ratio, where density increase and
shear deformation are observed simultaneously \shortcite{Rouxel2008}.

We would like to point out that the field of plasticity and rheology, even limited to amorphous materials, is a very broad one, 
and that the subject has been tackled by many groups with widely  different backgrounds, from metallurgy to soft matter and granular materials.
Such a brief review may in many cases give a very incomplete account of important theoretical developments. We hope however that the reader
will find in the bibliography the necessary references. In all cases, we tried to point out critically what we see as the  limitations
of theoretical approaches, in the hope that the reader's interest will be stimulated to test and possibly improve the current models.
Our bias in this chapter, consistently with the general topic of the book,  will be to put more emphasis on  models  that involve
 dynamically heterogeneous, collective behavior, and could  be analysed using tools presented in other chapters. This naturally 
leads us to insist on results obtained on systems in which thermal noise is not dominant (such as many soft matter systems, or glasses
at low temperature) and such heterogeneous behavior in the flow response is expected to be most important.

\section{Theoretical background}
\label{sec:2}

\subsection{Macroscale}

\begin{figure}
\begin{center}
\includegraphics[width=10cm]{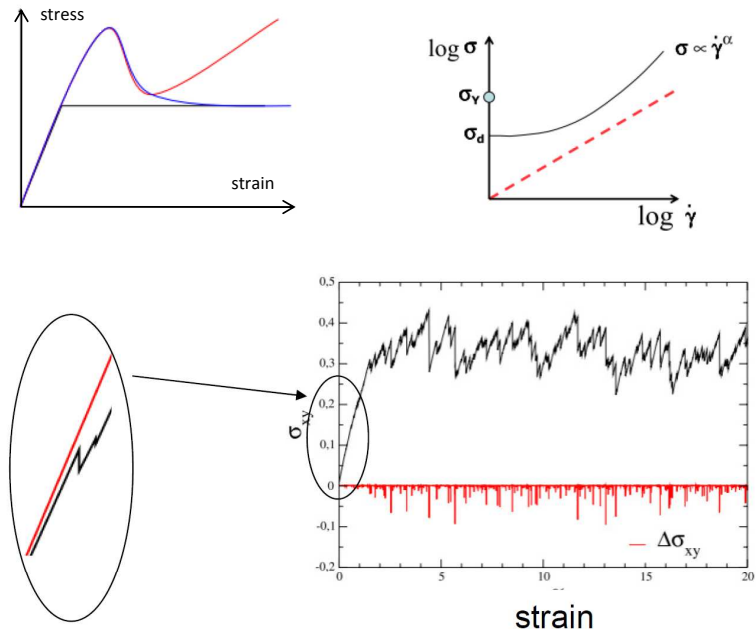}
\end{center}
\caption{Top left:  schematic illustration of a stress strain curve for the plastic deformation of a solid material. The three curves correspond respectively to ideal elastoplastic behavior (a), a metallic  (b) and a polymeric system at finite strain rate (c). The strain hardening part at large strain is specific to polymer systems. Top right: schematic illustration of a flow curve for a soft material; the dotted line is the Newtonian fluid case, while the low strain rate limit of the full curve corresponds to the yield stress. Bottom: actual response of a simulated strained glass at low temperature. Note the large stress fluctuations, which are associated with the finite size of the sample. The stress variations are also shown. The zoom on the elastic part at low strain shows that small plastic events are also observed in this part. Adapted from \protect\shortcite{TanguyLeonforteBarrat2006} }\label{fig:stress-strain}
\end{figure}

\paragraph{Plasticity at the macroscale}
Many tests have been developed to characterize plastic behavior. They involve deforming a piece of material under various conditions: constant stress, constant or oscillatory strain-rate, step stress or strain,... while measuring the response using appropriate observables. Typical stress responses, for a material loaded at constant strain-rate, are depicted in Figure~\ref{fig:stress-strain}. When the strain is vanishingly small, the material responds elastically. As strain increases, plasticity progressively sets in and the stress smoothly rounds off. This transient response obviously depends on the evolution of the internal state of the system during loading, hence on sample preparation and loading rate. Further increase of strain may be accompanied, in some circumstances, with the onset of an instability, as deformation localizes along shear bands and leads the material to failure. When this instability can be avoided, however, stress reaches a steady state plateau value.

In general, it is difficult experimentally to achieve situations in which the flow is truly uniform: external driving must always be applied at boundaries, so that material flow must be inhomogeneous to adapt to these conditions (one exception is the flow of a Newtonian fluid in plane Couette geometry). So, in the flow of complex fluids one often observes a strong nonuniformity 
in the flow rate which is due to the combination of a nonuniform stress field with a strongly nonlinear flow curve, and one often speaks of ``localisation'' in this context. This use of the  term  ``localisation'' is most often found in rheology, as typical experimental set-ups, such as the Couette cell necessarily lead to flow profiles which are inhomogeneous~\shortcite{Coussot2005}. ``Localisation'' in also used in another context, for example, in experiments on ``hard'' glasses~\shortcite{SchuhHufnagelRamamurty2007}, when it refers to an instability which is not directly associated with the macroscopic stress inhomogeneity and can only be observed during transients  as it leads to failure. Whether these two forms of localisation are related remains an open questions. 

\paragraph{Constitutive equations}

One goal of theories of plasticity is to provide a macroscopic description of the deforming medium, analogous to the Navier-Stokes equations for Newtonian fluids. As illustrated in Figure~\ref{fig:stress-strain}, plastic materials exhibit significant, preparation or age-dependent, transients in which stress does not match its steady state value. Proper account of either transient response or of instabilities of the steady inhomogeneous flow must rely on a description of how the internal state of the system depends on loading history. This entails identifying variables to properly characterize the material state and providing constitutive equations to complement the relevant conservation equations (momentum, but also possibly density or energy). The search for constitutive equations often rests on the idea that it is possible to provide a local and instantaneous representation of the material state, that is to describe its response in terms of relations between local quantities (such as stresses, strains, energy, density,...) and their derivatives.

The introduction of internal state variables is also a necessity if one wants to study localisation, or instabilities in the material response. For example, it has been  shown that constitutive equations which are approximated (grossly) by their steady state stress/strain-rate relation lead to instability criteria which are rudimentary, and that internal variables have to be introduced to capture non-stationary material response~\shortcite{RiceRuina1983}. Situations when the plastic response is unsteady play a critical role in the identification of relevant state variables. Localisation thus attracts a considerable amount of interest not only for its practical importance, but also for the theoretical implications of identifying the proper variables governing its development.

\paragraph{State variables for localisation}

One of the most obvious state variable is temperature. Clearly, it is expected to increase due to plastic activity when a material is strained under adiabatic conditions. Its role in localisation has thus been the subject of a long-standing debate in the metallic glass community~\shortcite{SchuhHufnagelRamamurty2007}. Some insist that localisation could be due to local heating and should be treated by introducing thermal dependence in the stress/strain rate relationship, plus energy conservation equations. The consensus, however, now is that this argument fails because the dissipated energy due to plasticity is evacuated too fast~\shortcite{SchuhHufnagelRamamurty2007}. Other features of material response, such as localisation or the peak stress, depend sensitively on age at time scales which are completely separated from those of thermal exchanges. These observations rule out the role of thermal inhomogeneities in most cases. In soft materials, the solvent acts in general as an external bath, so that deformation can be considered to take place under isothermal conditions.

A model proposed recently to account for shear banding or fracture focusses on compressibility and density fluctuations~\shortcite{FurukawaTanaka2006,FurukawaTanaka2009}. Like the thermal theories of localisation, it is based on a purely macroscopic description. It assumes that (i) the flow is Newtonian and (ii) stresses assume their steady-state values. The values of shear stress and pressure are thus provided by simple steady state equations, function of strain-rate and density, and localisation is in a sense a kinematic effect. The proposed  scenario involves a strong density dependence of the viscosity on density or pressure. As a result, a density fluctuation
(which is usually neglected in the description of incompressible flows) results in a local decrease of the viscosity, which in turn increases the local shear rate
and leads to an amplification of the fluctuation.
A linear stability analysis of the Navier Stokes equation under these conditions shows that for shear rates larger than $(\partial\eta /\partial P)^{-1}$,
large wavelength density fluctuations at 45 degrees from the flow direction become linearly unstable. This leads to growth of density fluctuations and
potentially to material failure.

Similar arguments have been used to describe the coupling between density fluctuations and rheology in polymer solutions~\shortcite{Helfand1989}.
This type of scenario predicts that strain localisation only operates {\it above} a critical strain rate, and it was argued in~\shortcite{FurukawaTanaka2009} that it captured a transition to localisation which is seen around $T_g$. Experiments on metallic glasses also show, at low temperatures, an enhancement of localisation at low strain rates, and a form of localisation which now occurs when the strain rate lies below a critical value~\shortcite{SchuhHufnagelRamamurty2007}. It thus seems that the localisation predicted in \shortcite{FurukawaTanaka2009} around $T_g$ is different from the phenomenon which is discussed in metallic glasses at low temperatures.

\paragraph{Discussion}

These difficulties illustrate the problems encountered when formulating theories of plasticity. It seems unlikely that a description involving only the usual thermo-mechanical observables (stress, pressure, temperature) can be sufficient. We know indeed that these variables do not provide a complete representation of the internal structure of glasses.  The state of a glassy system is, by definition, out of equilibrium, and evolves constantly with time in a relaxation process that 
 involves hopping in a potential energy landscape (PEL)~\shortcite{Stillinger1995,DebenedettiStillinger2001}, over distributed energy barriers~\shortcite{DoliwaHeuer2003,DoliwaHeuer2003a,DoliwaHeuer2003b}. Deformation modifies  the picture as it competes with this relaxation~\shortcite{UtzDebenedettiStillinger2000} and constantly rejuvenates the glassy structure, i.e. allows the system to stay in rather high energy states, and in some cases to reach a nonequilibrium steady state. The question is thus, using a few dynamical equations involving a limited set of state variables, to be able to describe the response of a glass driven out-of-equilibrium by external deformation. To describe the transient stress response upon loading, as exemplified in Fig.~\ref{fig:stress-strain}, we would also have to be able to describe the state of the glass which is produced by annealing in terms of variables that can be introduced in a description of plastic response.


The description of flow in amorphous solids that will be presented in the following is essentially based
on ideas from elasticity theory, and  therefore starts
from the low temperature, solid side of the glass or jamming transition. We mention briefly here another alternative approach,
the mode-coupling approach of Fuchs and Cates \shortcite{FuchsCates2002}.  Let us recall that the mode coupling
approach to dense liquids involves a nonlinear feedback mechanism, in which the relaxation dynamics
of a given density fluctuation at some  ${\bf k}$ is coupled to that of all other fluctuations at different wavevectors.
This feedback leads to structural arrest -- i.e. absence of relaxation and appearance of a frozen structure -- at some finite density and temperature.
The effect of shear described by Fuchs and Cates is the advection of density fluctuations, in such a way that the coupling
that leads to this non linear feedback weakens with time, and  a relaxation eventually results.  A dynamical yield stress
is obtained as the zero shear rate limit of the stress under steady shear, and turns out to be non zero
for the high density, strongly coupled systems that the theory predicts to be in a nonergodic state at zero shear.
The theory was developed further  by Brader and Fuchs \shortcite{BraderFuchs2008}, and predicts rheological behavior that is in very reasonable agreement with experimental observations on
colloidal glasses, together with a prediction of the deviation from the equilibrium fluctuation-dissipation theorem.  However, this mode  coupling description
 has not been extended, up to now,
to the description of dynamical heterogeneities.

\subsection{Local Inelastic Transformations}

In crystals, the mechanisms of plastic deformation involve the motion and generation of dislocations,
which are a class of topological defects of the periodic structure. The possibility to identify and precisely
define the objects responsible for irreversible deformation has played and continues to play a key role
in formulating phenomenologies and constructing governing equations for crystalline plasticity.
In contrast, the state of knowledge regarding amorphous systems lags far behind, because the elementary objects governing the plastic response remain quite difficult to pinpoint.

Most modern theories of plasticity are based on the idea, proposed by Ali Argon in the late 70's~\shortcite{Argon1979}, that macroscopic plastic deformation is the net  accumulation of local, collective, rearrangements of small volume elements--typically 5-10 particles in diameter. Initially corroborated by the observation of flow in a bubble raft~\shortcite{ArgonKuo1979}, this idea is now firmly supported by numerical simulations~\shortcite{ArgonBulatovMottSuter1995,FalkLanger1998,MaloneyLemaitre2004a,MaloneyLemaitre2004b,MaloneyLemaitre2006,TanguyLeonforteBarrat2006}, and by a recent experiment~\shortcite{SchallWeitzSpaepen2007}. 

In contrast with crystals, however, the notion of rearrangements (also called ``local inelastic transformations'', ``shear transformations'', or simply ``flips'') refers to a process, not to a specific type of defect.
Means have been devised to identify ``zones'' before they flip, via measurements of particle displacement fields,~\shortcite{LemaitreCaroli2007} or of local elastic moduli~\shortcite{Yoshimoto2004,Tsamados2009}, but these observations can be unequivocally correlated with rearrangements only rather close to yielding, and there is no  universally accepted prescription for
identifying  \emph{a priori} the locations (the ``zones'') where flips occur. Of course, it seems reasonable to infer that there should be some specific features of the  local packing, which make these transformations possible--and are probably related to the fluctuations of local thermodynamic quantities such as energy, stress, or density (free-volume). Many terms have been coined to reflect this notion--``flow defects'',~\shortcite{Spaepen1977} ``$\tau$-defects'', or more recently ``shear transformation zones'' (STZ)~\shortcite{FalkLanger1998}-- but the questions of what precisely a zone
is, or how zones could be identified before-the-fact, remain largely 
open.

This idea does not imply that only \emph{shear} transformations exist, or  that each rearrangement is a pure shear event. Rearrangements may involve some local changes of volume too, and some theories--such as free-volume approaches--may attempt to take this into account. Occurring at a very local scale, rearrangements are also inevitably broadly distributed. But in view of constructing a theory of stress relaxation, it is the net effect of local contributions to shear strain which clearly matter.
\begin{figure}
\begin{center}
\includegraphics[width=10cm]{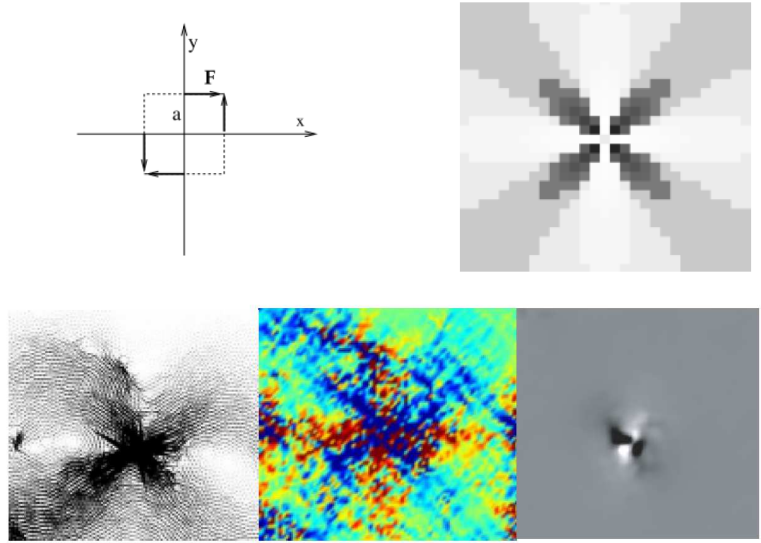}
\end{center}
\caption{Top: schematic illustration of a force quadrupole that corresponds to the theoretical description of shear transformation zones (left), and its Eshelby stress field. Bottom, from left to right: displacement field, stress response, and energy change associated with a localized plastic event, as observed in the quasistatic simulations described in section \ref{sec:particles}. Figures adapted from refs \protect\shortcite{PicardAjdariLequeuxBocquet2004,MaloneyLemaitre2004b,TanguyLeonforteBarrat2006}.}
\label{Eshelby}
\end{figure}

\subsection{Activation Theories}

Activation theories attempt to describe plasticity as the net result of independent shear transformations, which are supposed to be rare, thermally activated events. It is thus assumed that the system spends most of its time near local equilibria: clearly, this picture belongs to a low temperature regime, when the dynamics in a PEL would be described by infrequent hops between local minima. With these postulates, constitutive equations require a specification of transition rates (as a function of energy, density, stress) governing the rearrangements.

\paragraph{Eyring's theory}

The simplest form of such an activation theory of material flow dates back to the work of Eyring, and was initially intended to account for the viscous behaviour of liquids (above $T_g$). In Eyring's view, flow proceeds by the motion of single molecules into holes left open by neighbouring ones. This obviously misses the possibility that elementary events are in fact collective, but Eyring introduces several assumptions which are still quite fundamental to the field and useful to keep in mind as a reference.

He first assigns a typical value $E$ to the energy barrier that must be overcome to allow such hops--it is, in his view, the energy needed to create a hole--and observes that various types of hops are possible, which can either increase or reduce the macroscopic stress. He then restricts his description to two types of opposite moves, contributing  elementary strains of opposite signs $\pm\Delta\epsilon_0$, so that the macroscopic strain-rate takes the form:
\begin{equation}
\label{eqn:eyring:0}
\dot\gamma = \Delta\epsilon_0\,\left({\mathcal R}_+-{\mathcal R}_-\right)
\end{equation}
with ${\mathcal R}_\pm$ the rates of forward and backward moves. These rates are taken to follow an Arrhenius activation law describing $\pm$ hops over activations barriers:
\begin{equation}
\label{eqn:eyring:R}
{\mathcal R}_\pm= \omega_0\,\exp(-E_\pm/kT)
\end{equation}
with $\omega_0$ a microscopic frequency. Eyring assumes that  stress induces a linear bias between energy levels, so that the barrier depends linearly on $\sigma$: $E_\pm= E_0\mp \sigma\,\Omega_0$, with $\Omega_0$ an ``activation'' volume. This leads to writing the stress/strain-rate relation as:
\begin{equation}
\label{eqn:eyring}
\dot\gamma = 2\,\omega_0\,\Delta\epsilon_0\exp\left(-\frac{E_0}{k T}\right)\sinh\left(\frac{\Omega_0\sigma}{k T}\right)
\end{equation}
Of course, the linearisation of this relationship for small stresses leads to Newtonian behaviour, with an Arrhenius viscosity. At large $\sigma$, it is customary to keep only the dominant exponential term and write the stress as:
$$
\sigma = \frac{k T}{\Omega_0}\,\ln\left(\tau_0(T)\dot\gamma\right)
$$
with $\tau_0(T) = \exp\left(\frac{E_0}{k T}\right)/\omega_0\Delta\epsilon_0$.

\paragraph{Argon's theory}

Eyring's formulation relied on a representation of deformation in terms of hops of single atoms or molecules, instead of collective motions. Early representations of plasticity addressed this issue while borrowing from the general framework  of Eyring's approach, and in particular the idea that the linear viscous behaviour arises from a balance between forward (stress releasing) flips and a ``back flux''. In order to take into account the collective character of elementary transitions, Argon~\shortcite{Argon1979} argues that the shear zones can be viewed as inclusions which are elastically coupled to the surrounding medium, and postulates that a flip occurs when a zone elastically deforms up to some critical strain, in the range of $\sim2-4\%$, at which it becomes unstable. The calculation borrows from Eshelby's work on martensites~\shortcite{Eshelby1957}, which offers an analytical framework to estimate the total change in elastic energy due to a change in the internal strain state of an inclusion.

The question is how to take into account the fact that the average stress level biases the elastic energy associated with   a zone. At high temperatures, Argon computes the stress bias at linear order, like Eyring, leading to an expression similar to~(\ref{eqn:eyring}). At low temperatures, he performs an imposing treatment of the elastic problem, to arrive at a perturbative, second order, estimate of the effect of stress on the minimum where the system resides. It leads to an expression of the form~\shortcite{Argon1979}:
\begin{equation}
E_+ \propto \left(1-\frac{\sigma}{\sigma_c}\right)^2
\label{eqn:sigma2}
\end{equation}
where $\sigma_c=\mu(T)\epsilon_c$ is a typical scale of the shear stress needed to reach the strain  $\epsilon_c$ at the yield point and $\mu(T)$ is the shear modulus.  Using equations~(\ref{eqn:eyring:0}),~(\ref{eqn:eyring:R}) and neglecting the back-flux, equation~(\ref{eqn:sigma2}) leads to a stress/strain-rate relation of the form: $\sigma-\sigma_c\propto-\left(T\,\ln\left({\Delta\epsilon_0\,\omega_0}/{\dot\gamma}\right)\right)^{1/2}$.
The main interest of this theory is that it captures important features of experimental data on metallic glasses, in particular, (i) an apparent singular behaviour in the low-$T$ limit, and (ii) the weak $\dot\gamma$ dependence over a broader range of temperatures~\shortcite{SchuhHufnagelRamamurty2007}.

One main limitation of this approach is that it treats perturbatively--at second order--the effect of stress on the elastic potential of a shear transformation zone. This approach must break down if the stress level is sufficiently large to bring a zone close to instability, as the system then approaches a catastrophe. Indeed there is evidence from numerical simulations that deformation-induced instabilities correspond to a saddle node bifurcation~\shortcite{MalandroLacks1997,MalandroLacks1999,MaloneyLemaitre2004a}. The energy barrier near instability is thus of the form $E\propto\left(1-{\sigma}/{\sigma_c}\right)^{3/2}$ which, after insertion in equations~(\ref{eqn:eyring:0}) and~(\ref{eqn:eyring:R}) would lead to a stress/strain-rate relation of the form: $\sigma-\sigma_c\propto-\left(T\,\ln\left({\Delta\epsilon_0\,\omega_0}/{\dot\gamma}\right)\right)^{2/3}$.
Already formulated by Caroli and Nozi\`eres~\shortcite{CaroliNozieres1996} in the context of dry friction, this argument was brought up recently by Johnson and Samwer~\shortcite{JohnsonSamwer2005} for metallic glasses, and successfully compared with experimental data.

\paragraph{Discussion}

A first objection which should be made against these activation theories is that they treat flips as uncorrelated events. As Bulatov and Argon first noted~\shortcite{BulatovArgon1994a,BulatovArgon1994b,BulatovArgon1994c}, however, each rearrangement creates a long-range elastic field (Eshelby~\shortcite{Eshelby1957}), hence alters the stress in the rest of the system. These stress changes can be viewed as mechanical signals which are emitted by flips and may trigger secondary events~\shortcite{LemaitreCaroli2009}. This mechanism shows up strikingly in numerical simulation of athermal systems, via the emergence of avalanche behaviour~\shortcite{MaloneyLemaitre2004b,DemkowiczArgon2005,MaloneyLemaitre2006,BaileySchiotzLemaitreJacobsen2007,LernerProcaccia2009,LemaitreCaroli2009}. When these elastic effects are present, it becomes essential to understand the role of the self-generated stress noise in the activation of plastic events themselves.

A second objection is that the above theories obviously ignore the fact that the energy barriers which limit plastic transformation are broadly distributed~\shortcite{DoliwaHeuer2003,DoliwaHeuer2003a,DoliwaHeuer2003b}. Moreover, as the material is strained, shear transformation zones are driven towards their instability threshold~\shortcite{MaloneyLemaitre2004a,LemaitreCaroli2007}. This is quite different from the situation of a glass undergoing thermal relaxation: here, the distribution of barriers is set by the competition between elastic loading and plastic yielding~\shortcite{RodneySchuh2009a}, and some regions of space could, depending on the parameters, present temporarily, very low energy barriers. Rather than being set a priori, the relevant barrier distribution must thus result from a complex dynamical process.

Third, activation theories of plasticity ignore the role of fluctuations of local quantities such as free-volume, elastic constants, stresses, etc. As we will see in Section~\ref{sec:particles}, however, the idea that yielding can  easily be associated with a unique critical value of the local stress or energy  is strongly challenged by numerical observations \shortcite{TsamadosTanguyLeonforteBarrat2008}. Moreover, the activation process should depend significantly on details of the atomic packing, in particular via the fluctuations of local elastic moduli~\shortcite{Mayr2009,Tsamados2009}, local density or pressure levels~\shortcite{DemkowiczArgon2004,DemkowiczArgon2005,ArgonDemkowicz2006}.

\subsection{Dynamics of the local stress field}

A number of models~\shortcite{BulatovArgon1994a,BulatovArgon1994b,BulatovArgon1994c,BaretVandembroucqRoux2002,PicardAjdariBocquetLequeux2002,PicardAjdariLequeuxBocquet2004,PicardAjdariLequeuxBocquet2005} focus on the local stress field and its dynamics, in order to take into account two important features which were identified in the previous discussion: (i) the fact that loading drives the system towards local instability thresholds, hence that barriers result from a dynamical process, and (ii) the fact, that rearrangements may interact via their Eshelby stress fields.

The first of these models was proposed by Argon and Bulatov~\shortcite{BulatovArgon1994,BulatovArgon1994a,BulatovArgon1994b,BulatovArgon1994c} in a study involving various aspects of plasticity but also of glassy relaxation~\shortcite{BulatovArgon1994b}. They showed that even in the absence of deformation, taking into account long-range interactions between shear transformations could lead to net behaviour which ressembles that of glassy models incorporating e.g. distributions of timescales. Their model involves a collection of weak zones which are distributed on a triangular lattice. Transformation probabilities are determined from activation theory, with stress-dependent free-energy barriers $\Delta G^*(\sigma)=\Delta F_0-\Omega\sigma_{\alpha\beta}\Delta \epsilon_{\alpha\beta}$ where $\Delta \epsilon_{\alpha\beta}$ is the strain increment of a transforming zone during transition. Only pure shear transformations are allowed for convenience, but this does not exclude dilation which is introduced via a dilation parameter meant to account for how activation energy depends on pressure. Finally, each transformation alters the stress levels in the whole system via a Green tensor, corresponding to the solution of the Eshelby problem. The zone flips are thermally activated, but the barriers depend of the stress sustained by the zones, so that the introduction of this mechanism allows for correlations between flips.


The gist of Bulatov and Argon's viewpoint is that initially, after a quench, many of these weak zones should be far from their threshold as a consequence of annealing. As stress increases, energy barriers decrease, and some flips might occur. For small values of external stresses, only a few zones, sufficiently near their instability thresholds at the end of annealing, will respond. Moreover, as most zones are far from instability, the stress released by these rare flip events is insufficient to trigger secondary events. Initial loading is thus accompanied by a small number of isolated rearrangements. As stress increases, however, an increasing number of zones come close to their instability thresholds. Mechanical noise then starts to be able to trigger events elsewhere, leading in some cases to localisation.

Two more recent models, strongly inspired by the Argon Bulatov model, have been proposed by Baret and coworkers \shortcite{BaretVandembroucqRoux2002} and Picard and coworkers \shortcite{PicardAjdariBocquetLequeux2002,PicardAjdariLequeuxBocquet2004,PicardAjdariLequeuxBocquet2005}. The model of reference \shortcite{BaretVandembroucqRoux2002} has the interesting feature of incorporating a distribution of threshold values for the local yield stress, but uses extremal dynamics to describe the evolution of the system at vanishingly small strain rate. The model of Picard, on the other hand, does not include disorder, but has the advantage of relative simplicity and of easily incorporating finite strain rate effects. The model describes the evolution of a scalar stress $\sigma_i$ on a lattice site via an equation of the form
\begin{equation}\label{picard1}
\partial_t  \sigma_i = \mu \dot{\gamma} + \sum_j G_{ij} \dot{\epsilon}_{j,\rm plast}
\end{equation}
where $\mu$ is a shear elastic modulus. The first term describes an elastic loading proportional to an average external strain rate. The second term is the supplementary loading that arises from the plastic activity at all other sites in the system, which is assumed to be transferred instantaneously through an elastic propagator $G_{ij}$. This plastic activity $ \dot{\epsilon}_{j,\rm plast} $ is computed in turn by assuming that any site that reaches a stress beyond some local critical yield $\sigma_Y$ releases its stress with a time constant $\tau$. Numerical studies of the model show that, at low strain rates, zones of persistent plastic activities can be observed, with a typical size that tends to diverge as the strain rate vanishes.


\subsection{Dynamics of distributions}

Based on the ideas mentioned above, several approaches have been proposed that describe the state of the system by the distribution of a scalar variable--corresponding to either stress levels~\shortcite{HebraudLequeux1998} or energy barriers~\shortcite{SollichLequeuxHebraudCates1997,Sollich1998}.
They are based on some empirical assumptions that are inspired by the physical picture
of interacting flow defects. In most of these approaches, an assumption of homogeneous deformation  is made, so that the description of macroscopic strain localisation is not permitted. Moreover, a scalar description of stress and strain is retained,
with the implicit assumption that the scalar stress corresponds to a local shear stress. Despite this apparent simplicity, these
models are far richer that simple local constitutive equations, as they introduce in the picture some auxiliary quantity describing the internal state of the system, in the spirit of the "rate and state" models of solid friction \shortcite{RiceRuina1983}\footnote{Simpler rate and state models involving a single "fluidity" internal parameter have also been proposed in refs.~\shortcite{PicardAjdariBocquetLequeux2002,CoussotNguyenHuynhBonn2002}.}.

\paragraph{H\'ebraud-Lequeux fluidity model}
The simplest of these models is probably the one introduced by H\'ebraud and Lequeux \shortcite{HebraudLequeux1998}.
In this model, one
deals with an ensemble of sites that can each sustain a stress $\sigma$. The central quantity is the probability distribution function (pdf)
$P(\sigma,t)$ of the local  stress, which is assumed to evolve according to the equation
\begin{eqnarray}\label{hl1}
\partial_t  P(\sigma,t) = &  -G_0 \dot{\gamma} \partial_\sigma  P(\sigma,t)
-\frac{1}{\tau}H(|\sigma|-\sigma_c) P(\sigma,t) 
\cr
 +&
\frac{1}{\tau}  \delta(\sigma) \int_{|\sigma'| >\sigma_c} P(\sigma',t) d\sigma'
+ D \partial^2_{\sigma^2}P(\sigma,t)
\end{eqnarray}
where $H$ is the Heaviside step function, and  the "stress diffusion constant" $D$ is given self consistently by
\begin{equation}\label{hl2}
D= \frac{\alpha}{\tau}  \int_{|\sigma'| >\sigma_c} P(\sigma',t) d\sigma'
\end{equation}
Equation (\ref{hl1}) is a simple evolution equation for the pdf: the first term correspond to elastic loading at constant strain rate,
with  an elastic modulus $G_0$. The second and third terms correspond to a description of plastic events that take place with a rate $1/\tau$
for sites that exceed the critical yield stress $\sigma_c$; according to the third term, each plastic event corresponds to a complete release of the stress,
which is set equal to zero.
The last term describes a "diffusion" along the stress scale, that is the result of the average activity
(stress redistribution after loading or unloading)  of all other sites. This could also be described as a "stress noise", and the intensity of this
noise is taken, according to equation (\ref{hl2}), to be proportional to the total plastic activity present in the third term of equation \ref{hl1}.
The coupling parameter $\alpha$ in eq. (\ref{hl2}) is the control parameter of the model. It could be interpreted as
 corresponding  to the intensity of the elastic coupling between sites. For small values of $\alpha$ ($\alpha<\alpha_c$) the system is
 jammed, with a vanishing activity $D=0$ in the absence of  strain, and multiple solutions for $P(\sigma)$.  In this jammed
 situation, the model exhibits a nonzero yield stress (the limit of $\langle \sigma \rangle$ when $\dot{\gamma}$ goes to zero is nonzero)
 and a complex rheological behaviour. Despite its apparent simplicity, this model illustrates how the introduction of couplings
 between simple elasto-plastic elements, even when treated at the mean field level, can give rise to a complex collective behaviour.
 Such models can serve as a basis for more complex, non-local fluidity models, as discussed below.

\paragraph{Soft Glassy Rheology}

The first, very successful example of a model using the dynamics of a distribution function is probably the "Soft Glassy Rheology" approach of
Sollich and coworkers~\shortcite{SollichLequeuxHebraudCates1997,Sollich1998}. A number of reviews of this approach, which has been very successfully applied to many features of the rheology of soft glasses, including complex strain histories, are available \shortcite{Cates2002}. We will therefore limit ourselves to a very brief description of the assumptions underlying this model. The  system is a collection of independent elastoplastic elements, each of which is trapped in an energy minimum of depth $E<0$ (relative to some zero energy level). Each element is also assigned a strain $\ell$ that 
varies with time as the global strain $\gamma$, and the energy barrier changes with strain as $E\rightarrow E(\ell)=E+k\ell^2/2$, where $k$ is a local modulus. The escape from a potential well (corresponding to the local yield of an element) is governed by an Arrhenius-like factor, $\tau_{\rm yield} \sim \exp(-E(\ell)/X )$. The two key
ingredients in the model are (i) a distribution of trap depths that implies, in the absence of external strain and for small values of $X$, that the system has a very broad distribution of relaxation times, and is effectively in a glassy state, with a "weak ergodicity breaking" described by the trap model of Bouchaud \shortcite{Bouchaud1992} ; (ii) the "effective temperature parameter" $X$, which activates the dynamics of any given element, and is intended to represent the mechanical noise arising from the yield of all other elements in the sample. The introduction of $X$
is a recognition that, in many systems, thermal motion alone is not enough to trigger local yield events. The system has to cross energy barriers that are very large compared to typical thermal energies. However, the model suffers from the fact that this  parameter is not determined self consistently, and therefore should be considered as adjustable.  The H\'ebraud-Lequeux equations described above can be considered as a first, simplified  attempt to obtain self consistency within this type of framework.

\subsection{Classical rate-and-state formulations}

The models we have described so far rely on the notion of zone flips in a weak sense: they assume macroscopic plastic behaviour to be the net effect of sudden local rearrangements which can occur anywhere within the material depending primarily on the local or macroscopic stress level; but their focus is on the processes--the flips--but not on their loci--the zones. Underlying all of these theories is of course the notion that some regions of space present specific traits--low density? modulus?--which facilitate flips. Rate-and-state formulations attempt to incorporate dynamical equations governing the density of the ``flow defects''~\shortcite{Spaepen1977} or ``shear transformation zones''~\shortcite{FalkLanger1998}, which are supposed to control plastic activity.

\paragraph{Free-volume}

The introduction of free-volume dynamics in rate-and-state formulation of plasticity was pioneered by Spaepen as early as the 70's. It borrows from Cohen and Turnbull's free-volume theory~\shortcite{CohenTurnbull1970}, which was proposed to explain departures from Arrhenius behaviour by introducing a variable which would ``replace'' temperature in activation factors. A material is seen as decomposed into many sub-systems of local free-volume $v_i$. These variables are then assumed to be exponentially distributed (with little justification) and relaxation events are assumed to occur in regions of high free-volume $v_i>v_0$, with density $\propto e^{-v_0/v_f}$ (where $v_f=\langle v_i \rangle$ is the average free volume).
Spaepen then argues, that the strain-rate must then take the form~\shortcite{Spaepen1977,HeggenSpaepenFeuerbacher2005}:
\begin{equation}
\dot\gamma \propto \Delta\epsilon_0\,\omega_0\,e^{-v_0/v_f}
\,\sinh\left(\frac{\Delta\epsilon_0v_0\sigma}{2kT}\right)
\end{equation}
To couple the plastic response with changes of density, he proposes to write an evolution equation for the concentration $c_f=e^{-v_0/v_f}$ of high free-volume defects, in the form:
\begin{equation}
\dot c_f = -k_rc_f(c_f-c_{\rm eq})
\end{equation}
with a constants $k_r$ and $c_{\rm eq}$, which depend on temperature and stress. This equation captures the idea that the evolution of $c_f$ is controlled by two antagonistic effets, (i) deformation introduces dilatancy at a rate supposedly proportional to $\dot\gamma$ and (ii) high free-volume defects occasionally collapse via a ``bimolecular'' process--whence the form $c_f^2$,. . The competition between these two effects
determines an equilibrium value $c_{\rm eq}$ in steady state.

In this argument, the postulate that density fluctuation are exponentially distributed, or the precise form of free-volume creation and destruction terms should of course be questioned--but the same remark can be made regarding the creation/destruction term $\mathcal{X}$ in STZ theory (see further, equation~(\ref{eqn:stz:1})). A particular oddity in Spaepen's argument is the reference to a ``bimolecular'' process to explain the disappearance of free-volume, as if high-density defects where actual objects, which were also mobile and likely to collapse whenever they meet. Different free-volume equations were proposed later~\shortcite{Lemaitre2002b,LemaitreCarlson2004}, with a different interpretation, in particular, of the destruction process. It is based on the simple remark that if $v_f$ controls activation mechanisms and evolves in time, then its decay rate should be of the form $e^{-v_1/v_f}$, with $v_1$ an activation volume for the decay process which has no reason to be equal to $v_0$. The rate of dilatancy is moreover assumed to be proportional to $\sigma\dot\gamma$. Introducing the notation $\chi=v_0/v_f$, this leads to coupled equations of the form~\shortcite{Lemaitre2002b,LemaitreCarlson2004,Lemaitre2007},
\begin{eqnarray*}
\label{eqn:chi}
\dot\gamma &=& \Delta\epsilon_0\,\omega_0\,e^{-1/\chi}\,\sinh\left(\frac{\sigma}{\mu}\right)\\
\dot\chi &=& -A_0\,e^{-\kappa/\chi}+\sigma\,\dot\gamma
\end{eqnarray*}
which can reproduce various forms of the transient and unsteady response of amorphous materials under shear.

The free-volume parading associates (via $e^{-v_1/v_f}$ factors) large changes in relaxation timescales to minute changes of density. Precise analysis of experimental data for different conditions of density and pressure have shown, however, that is was unlikely the relaxation timescales could be thus governed by the amount of free-volume~\shortcite{AlbaKivelsonTarjus2002,KhonikKaverinKobelevNguyenLysenkoYazvitskyKhonik2008}. The existence of free-volume changes in plastic flow is also in question, as there is experimental evidence of homogeneous flows in which plasticity does not produce significant free-volume~\shortcite{HeggenSpaepenFeuerbacher2005}, while dilatancy effects have been observed during the formation of shear-bands~\shortcite{ChenChuang1975,DonovanStobbs1981,LiSpaepenHufnagel2002}. This observation, of course, does not mean that dilatancy actually controls the formation of shear bands, as it may only be a passive marker. We finally note that if the free-volume paradigm is not supported by experimental data for metallic or molecular glasses, it still could be relevant to other systems such as granular materials or colloidals glasses.

\paragraph{Dynamics of STZ densities}

The STZ model proposed by Falk and Langer~\shortcite{FalkLanger1998}, departs from prior works in its stricter interpretation of the concept of zones. They are seen as sufficiently well-defined objects which pre-exist the occurrence of flips, so that it actually makes sense to speak of their density, which becomes a dynamical state variable, somewhat analogous to a defect density in models of crystalline plasticity. Hence, a constitutive law for amorphous solids must include equations of motion for the density and internal state of these zones. Although recent papers have introduced tensorial formulations~\shortcite{Langer2004}, the gist of STZ theory is captured by assuming that zones are two-level ($\pm$) systems, which can produce strain changes $\pm\Delta\epsilon_0$ only when they undergo $\pm\to\mp$ internal transitions. This leads to writing an equation \`a la Eyring (see~(\ref{eqn:eyring:0})) for strain-rate as:
\begin{equation}
\label{eqn:stz:0}
\dot\gamma = \Delta\epsilon_0\,\left({\mathcal R}_+\,n_+-{\mathcal R}_-\,n_-\right)
\end{equation}
where $n_\pm$ stand for the number densities of $\pm$ zones. The dynamical equations for the number densities $n_\pm$ then take  the form:
\begin{equation}
\label{eqn:stz:1}
\dot n_\pm= -{\mathcal R}_\pm\,n_\pm+{\mathcal R_\mp}\,n_\mp+\mathcal{X}
\end{equation}
where the last term accounts for creation/destruction mechanisms which are meant to describe how the mean flow affects zone populations--in practice, it is assumed to depend e.g. on the macroscopic energy dissipation rate. The rate factors, of course, must then be expressed in terms of parameters such as stress, or density. In particular, a linear dependence in terms of stress will commonly be assumed.

Without going into any further details, it is already possible to see
from equation~(\ref{eqn:stz:0}) that there are two ways jamming can be
explained within the context of STZ theory. Indeed, the strain-rate
vanishes when ${\mathcal R}_+\,n_+={\mathcal R_-}\,n_-$, which may occur either
as the result of a balance between $\pm$ flips (in which case ${\mathcal
R}_+\,n_+\ne0$), or because the effective rate of flip events ${\mathcal
R}_+\,n_+={\mathcal R_-}\,n_-=0$.

Early STZ
papers~\shortcite{FalkLanger1998,Langer2001,EastgateLangerPechenik2003,LangerPechenik2003}
proposed that jamming occurred primarily via a ``polarization'' of the
medium, the jammed state being when ${\mathcal R}_+\,n_+={\mathcal
R_-}\,n_-\ne0$, which occurs for a specific value of the ratio
$n_+/n_-={\mathcal R}_-/{\mathcal R}_+$. The possibility to model jamming in this
manner is a direct consequence of the representation of zones as
two-level systems, which is also the particularity of STZ theory. Under
various assumptions for the transition probabilities and for the
creation/destruction term $\mathcal{X}$ in~(\ref{eqn:stz:1}), the STZ
equations were shown in~\shortcite{FalkLanger1998} to present a steady
plastic flow and a transition from jammed to flow at a limit external
stress, in a way which is consistent with expectations.

Later works on STZ theory~\shortcite{Lemaitre2002b,LemaitreCarlson2004,FalkLangerPechenik2004,BouchbinderLangerProcaccia2007,BouchbinderLanger2009,BouchbinderLanger2009b} introduce additional kinetic equations for a state variable $\chi$, called free-volume in~\shortcite{Lemaitre2002b,LemaitreCarlson2004} and ``effective temperature'' in~\shortcite{FalkLangerPechenik2004,Lemaitre2007,BouchbinderLangerProcaccia2007,BouchbinderLanger2009,BouchbinderLanger2009b},
such that $\exp(-1/\chi)$ accounts for a density of ``flow defects''.
The effective temperature is thought as a measure of the degree of
disorder in the system, and its dynamics is supposed to result from the
competition between plastic work--which drives the system more
out-of-equilibrium--and the spontaneous relaxation of the system towards
equilibrium.
  The resulting coupled equations were shown to be able to reproduce
various
aspects of the out-of-equilibrium response of glassy systems such as the
Kovacs effect~\shortcite{Lemaitre2007,BouchbinderLanger2009} or the formation
of shear bands~\shortcite{ShiKatzLiFalk2007,ManningLangerCarlson2007}.

As mentioned above, an  important  originality of STZ theory
-in particular compared to free volume theories that have
a very similar mathematical structure- lies  in the notion that zones
have an internal structure (modelled by the $\pm$ states). The physical
   motivation behind this idea was very clear in early
works~\shortcite{FalkLanger1998,Langer2001,EastgateLangerPechenik2003,LangerPechenik2003}:
it permits to explain jamming by a balance between forward and backward
rearrangements (${\mathcal R}_+\,n_+={\mathcal R_-}\,n_-$). This however implies
that in a jammed system, the plastic activity ${\mathcal R}_+\,n_++{\mathcal
R_-}\,n_-$ remains non-zero. To solve this problem for athermal systems, Falk and Langer have
introduced rates ${\mathcal R}_+$ and ${\mathcal R_-}\,n_-$- which are exactly
zero when the stress is applied opposite to the preferred direction of
the STZ.  In that case jamming  could come about from an exhaustion of
plastic activity, as expected in the low-$T$ range.
The introduction of the effective temperature also smooths out the
jamming mechanism with a crossover between the backward-forward
 balance and exhaustion, while  the two level aspect remains
needed to account for some specific features such as the
Bauschinger effect~\shortcite{FalkLanger2010}.

Despite its success in accounting for the macroscopic
mechanical behavior of metallic glasses, questions remain about some
basic assumptions of the STZ model. The very
definition of STZ's as ``ephemeral, noise activated, configurational
fluctuations that happen to be susceptible to stress-driven shear
transformation'', as presented in a recent review that covers the subject
extensively~\shortcite{FalkLanger2010}, makes a direct identification of
these zones (e.g. using numerical simulations) quite difficult. The same
can be said of the definition of the effective temperature, which, as its free
volume counterpart, does not appear to be directly observable
(although it could be related to potential
energy density in the simulations of ref.~\shortcite{ShiFalk2005,FalkLanger2010}). Consequently,
like in all the above-mentioned theories,  assumptions have to
be introduced concerning the form of activation rates, and how they
depend on stress, temperature, or  effective temperature.
Progress  and establishing
  more precise and microscopic definitions of
  zones or effective temperature, and understanding
these assumptions from a microscopic standpoint, will certainly stimulate future work
 and may lead to convergence
 with some of the alternative viewpoints such as SGR or fluctuation
 dissipation approaches \shortcite{HaxtonLiu2007}.

\subsection{Nonlocal rheology approach}

All the approaches discussed in the previous sections take for granted that some kind of microscopic local heterogeneity
governs the stress strain relationship in amorphous materials. As we will see below, this aspect has, in essence, been confirmed by particle based simulations, although the precise characterization of the flow events is still the object of numerous studies.   The next question is therefore the description of the spatial organisation of these events, both in terms of large scale fluctuations and of permanent strain localisation. The study of large scale fluctuations is a subject of current interest within particle based or lattice models~\shortcite{YamamotoOnuki1998,MaloneyLemaitre2004b,TanguyLeonforteBarrat2006,MaloneyLemaitre2006,BaileySchiotzLemaitreJacobsen2007,LernerProcaccia2009,LemaitreCaroli2009}, and some aspects will be discussed in section \ref{sec:particles}. Here we would like to discuss recent extensions of the mean field theories described above, that attempt a description of flow heterogeneities based on deterministic partial differential equations. The common point in all the recently proposed approaches is to  
extend the mean-field rheological
description by allowing for spatial variation of one of the parameters, based on a diffusive kernel~\shortcite{ManningLangerCarlson2007,GoyonColinOvarlezAjdariBocquet2008,FieldingCatesSollich2009}. For example, in \shortcite{ManningLangerCarlson2007}, the effective temperature parameter $\chi$ is assumed to obey  a diffusion equation, with a diffusion constant that is proportional to the local rate of plastic deformation.  In \shortcite{FieldingCatesSollich2009}, the effective temperature parameter $X$ of the SGR model obeys a relaxation-diffusion equation, with a source term that can be taken to be either proportional to the rate of local plastic activity or to the rate of energy dissipation. Finally, in \shortcite{GoyonColinOvarlezAjdariBocquet2008}, it is the fluidity--that is the local value of $\dot\gamma/\sigma$-- that appears inside the diffusive term. All these models have   been shown to reproduce various forms of localisation, either aiming to model the formation of shear bands in metallic glasses~\shortcite{ManningLangerCarlson2007}, or near-wall localisation in microfluidics flows of colloidal suspensions~\shortcite{GoyonColinOvarlezAjdariBocquet2008}, or even faults~\shortcite{DaubCarlson2008}.

This very important issue currently attracts considerable interest, but it seems from the diversity of the study, and their various claims to actually account for localisation, that many forms of diffusive kernels would capture some of the underlying physics, irrespectively of their actual microscopic assumptions. We must hence worry about the severity of tests provided by fitting such lengthscales. Clearly, it seems important to assess how sensitive the resulting behaviour is to particular forms of the diffusive terms, in other words, to break down the frontiers between different groups where these various equations are developed, and put them alongside each other for close comparison.

At the microscopic level, the specific mechanisms
which justify diffusive terms still have to be specified. At
present the only theoretical attempt to do this 
on the basis of a more microscopic approach can be found in ref.~\shortcite{BocquetColinAjdari2009}. This reference introduces a systematic coarse graining approach of the lattice elasto-plastic model of ref.~\shortcite{PicardAjdariLequeuxBocquet2004}, which eventually results in a coupling between a local rheology described by the H\'ebraud-Lequeux model with equation~(\ref{hl2}) being replaced by a nonlocal relation between stress diffusivity and the rate of plastic activity. In other words,
the stress diffusivity--in stress-space--is caused by the plastic activity in the vicinity of the zone in question--in real space. The Laplacian form of this contribution can be easily understood, as a purely linear gradient in plastic activity leads to a local compensation between neighbouring zones with higher and lower activities.

These approaches  are particularly attractive and will be amenable to a direct comparison with numerical data obtained from elasto-plastic lattice models or from particle based simulations, and probably with experiments. Such comparisons will be, in a first stage, based on the predictions for strain localisation and the characteristic length/times over which this phenomenon is predicted to take place. Further refinements, e.g. including tensorial aspects for the stress tensor, will be necessary to actually predict mechanical response under various perturbations.

\subsection{Conclusion}

\paragraph{What effective temperature?}

It is clear from the preceding discussion that rate-and-state formulations of plasticity should be seen as various forms of empiricism, rather than actual microscopic theories. This does not, however, diminish the fact that they do reproduce many features of experimental data. This is particularly striking in view of the simplicity of most of these formulations which do not involve complex, distribution-like state variables--like H\'ebraud-Lequeux or SGR. And of course, rate-and-state equations are potentially quite important as they can easily be incorporated in Navier-Stokes equations, and used to study flows with complex geometries.

It is striking that, taken as an empirical attempt to replace temperature in activation factors by another arbitrary quantity, the introduction of free-volume parallels--and anticipates~\shortcite{Spaepen1977}--the recourse to the notion of ``effective temperature''~\shortcite{CugliandoloKurchanPeliti1997,BerthierBarratKurchan1999,BarratBerthier2001,BerthierBarrat2002,OnoO'HernDurianLangerLiuNagel2002,HaxtonLiu2007} which seems more acceptable in today's language. This emphasizes the necessity to understand low-temperature activation mechanisms. Clarifying whether they involve stress noise, local fluctuations of pressure, density--like the free-volume theories would argue--or even moduli, however, remain open issues.

Free-volume models have the merit to make a specific, testable~\shortcite{Spaepen1977,HeggenSpaepenFeuerbacher2005}, 
assumption about the relation of activation factors with density. Sollich argues~\shortcite{SollichLequeuxHebraudCates1997,Sollich1998},
 that the effective temperature should represent the noise produced by ongoing rearrangements. Yet, as long as the link between
  ``noise temperature'' and defect densities is not established, it seems that these two different families of theories introduce
   the concept to take into account quite different forms of disorder. In fact, both mechanical noise and local fluctuations in
    moduli are present~\shortcite{Yoshimoto2004}, and both are related to the underlying distribution of local barrier height which, as
     we have discussed previously is determined by a dynamic interplay between elastic loading and plastic activity itself. In this regard,
      it is significant that the STZ and free-volume models require the introduction of  dynamical equations for their effective temperature. 

\paragraph{What zones?}

The notion of a shear transformation is  inevitably rather fuzzy as, in practice, rearrangements are hard to identify and isolate in space or time. However, it offers a useful 
framework for rationalizing observations,
and articulating
 theories. Phenomenologies of plasticity based on this premise must still specify a number of assumptions about flips, mainly in order to answer two questions: what triggers a flip? how much does a flip contribute to stress relaxation and energy dissipation?

These questions, at the core of the construction of theories, have motivated a number of  numerical and experimental studies, and some results will be discussed in section \ref{sec:particles}. The aim is to  attempt to check the basic assumptions contained in the theoretical
description, and to identify  the parameters which should be incorporated into theories.  This is of course difficult because zones are not easily identified within the disordered structure, and because material properties, such as elastic moduli strongly fluctuate.  Basically, it can be said that a consensus is now established on the
fact that elementary flips can be identified at low temperature, and produce a stress release characterized by a quadrupolar symmetry in two dimensions. Many questions, regarding in particular the conditions that trigger a flip, or the distribution of stress drops, remain to be explored accurately. The situation in three dimensions is even more complex, and even the form of the  stress field released after one flip has not been studied in any detail. The interaction between flips, the way a local rearrangement changes the probability for a subsequent flip to take place nearby, and the associated time scales, are also subjects of current interest which should be clarified and used as inputs to a theoretical description.

\section{Particle based simulations}
\label{sec:particles}

\subsection{Introduction}

Particle based simulations, and in particular molecular dynamics (MD) simulations
of systems under shear have a long history, as they have
been used very early to determine the viscosity of simple fluids.
A number of tools have been developed in this context that allow one to integrate the equations of motion
for an ensemble of a several thousands of interacting particles  under conditions of constant (or possibly oscillatory)
strain rate, or sometimes to conditions of constant stress.  Possible local heating associated with the
shear is taken care of using thermostats that do not perturb the shear flow (except possibly at very high shear rates which
will not be discussed in this paper).  Another "trick" that    allows one to mimic shear  in large systems
is the use of Lees-Edwards boundary conditions, in which the usual square (cubic) box periodicity is replaced by
that of a tilted Bravais lattice, with a tilt angle corresponding to a deformation  that  increases linearly with time. As a result the
system undergoes a shear deformation driven by the boundaries of the simulation cell. In some cases these Lees-Edwards
boundary conditions will be supplemented by a procedure that implements a homogeneous local deformation in parallel with the boundary driven deformation. In quasistatic deformations (see below), this is achieved by rescaling the particle coordinates at each strain step $\delta \gamma$ , in an affine manner e.g. $X^\prime= X+ Y\times \delta \gamma$. In MD simulations an algorithm called the SLLOD algorithm  \shortcite{AllenTildesley1996}  is used
to modify in a similar way the evolution of the velocity along the shear direction $x$,  $dv_x/dt = F_x/m + \dot{\gamma} v_y $, so that a linear velocity
 profile is immediately obtained after application of shear.

In the context of glassy systems, the shear rate is of course an essential quantity that provides a natural time scales in systems in which the
relaxation time is essentially infinite.  Particle based simulations are rather limited in terms of the time scale, with simulation times that are
at best of the order of   $10^6$ microscopic vibrational periods.
If within times of this order one wants to simulate total strains
$\gamma$ of order unity or higher, the resulting shear rates may seem
very high. For
a metallic system, they would be of the order of $10^8 s^{-1}$. Still, the   important feature is that a reasonable scale separation between the microscopic,
vibrational time scale and the  time scale of the deformation is achieved. A common assumption, confirmed to  some extent by results described below, is that this time scale separation is sufficient to make such simulations representative of the qualitative  behaviour in deformed glassy systems,
in spite of the very high rates used.

An alternative to these finite shear rate calculations is to use zero temperature,
quasistatic simulations. In these simulations the system is always in a local energy minimum.
After each elementary deformation step, which is described by an affine transformation of the particle coordinates,
 the energy is minimized to the "nearest" minimum using a conjugate gradient or steepest descent algorithm.
 Note that this "nearest minimum" may, strictly speaking,  depend on the minimisation algorithm.
 Here the notion of time step and duration of the simulation is totally absent, and is replaced by that of
 elementary strain step and of total deformation at the end of the simulation. The elementary deformation step is limited by the need
 to avoid an artificial "tunneling" through energy barriers during the elementary deformation. While this cannot be, strictly speaking, avoided,
 experience shows that elementary step strains of the order of $10^{-5}$  are small enough to limit such problems and to produce reproducible trajectories.
 As each step strain requires a careful minimization,  achieving large deformations with the quasistatic method is computationally costly. However, the method
 has the advantage of providing a well defined, "zero shear rate, zero temperature" limit for the flow behaviour of the system.

 A critical issue in the study of glassy systems using simulations is that of equilibration. The properties of a glass are known to depend
 strongly on the preparation route and on the quenching rate, which is un-physically high in simulations. The mechanical properties under
  small deformations are expected to be affected by the preparation history; however, if a large and homogeneous deformation can be achieved,
 the memory of the initial configuration will be erased, and the results will not depend on the preparation route. Note that the hypothesis of
 a  deformation that remains homogeneous at large scales is important here, and that different preparation routes may lead to samples
 that display a stronger tendency to strain localisation, without ever undergoing the homogeneous deformation that would erase its memory. Such
 effects were described in a model glass with partial quasicrystalline order in ref. \shortcite{ShiFalk2006}.

 Many different systems have been explored under shear using either quasistatic or finite shear rate simulations,
 periodic (Lees Edwards) cells or driving by external walls.  The studies published in the literature vary considerably in their
 choices for interaction potentials
 and also in the type of quantities that are characterized in the flowing systems.  Broadly speaking, one may distinguish three types of interaction potentials:
 mixtures of particles interacting by Lennard-Jones type potentials, that are generally used with the intention of modeling
 either metallic glasses \shortcite{SrolovitzMaedaEgami1981},  or colloidal suspensions \shortcite{StevensRobbins1993}.  Interactions with strong directional
 bonding are suitable for systems such as amorphous silicon or silica \shortcite{ArgonDemkowicz2006}, and  contact interactions are
 used to model granular systems \shortcite{CombeRoux2000} or foams \shortcite{Durian1995}\shortcite{LangerLiu1997}.  Polymers have also been extensively simulated \shortcite{ArgonBulatovMottSuter1995},  in general using Lennard-Jones type interactions with extra intramolecular bonding that gives rise to strain hardening at large strain under traction.

Particle based simulations offer the possibility to obtain, within the time and length scales permitted by simulation,
all microscopic information that can be obtained from particle coordinates. As often, the main difficulty is to find the
 appropriate tools to analyse the 
 large amount of data that is available, in
order to   extract the information relevant to the
 flow process and its heterogeneity.   In  the following, we describe some of the generic aspects that emerge
from these studies, without attempting an exhaustive literature review. We will distinguish results obtained at relatively high $T$,
where the thermal  fluctuations are important compared to those associated with the deformation, and 
results for low temperature, where microscopic motion originates essentially from the external driving.

\subsection{Finite temperature MD}

At finite temperatures, the local motion of particles is a complex superposition of thermal
and of deformation induced movements. The identification of specific plastic events associated
with the deformation is not possible. In practice, this situation is encountered if the temperature is close
to the glass transition temperature $T_g$, so that the systems at rest usually display significant ageing.
The  appropriate tools for analysing this regime   are largely inherited from
studies of the liquid-glass transition. A number of studies have focused on the global (macroscopic) stress strain relation,
and the occurrence of macroscopic, shear banding instabilities.
In spite of the short time scales explored by simulations, stress strain curves exhibit a behaviour that is remarkably similar to the
one observed in experiments. The peak stress in the curves shown in figure \ref{fig:stress-strain}  depends on shear rate and on
the "age" of the system, i.e. on the preparation history
\shortcite{VarnikBocquetBarrat2004}. This dependence was rationalized by R\"ottler and Robbins \shortcite{RottlerRobbins2005}
on the basis of "rate and state" ideas borrowed from solid friction. The peak stress varies as a logarithmic function of the strain rate
and of the age of the system
\begin{equation}\label{eq-peakstress}
    \sigma_{max} =  \sigma_0 + s_0 \ln \theta + s_1 \ln \dot{\gamma}
\end{equation}
where $\theta$ is an effective age of the system, that depends on the waiting time after the quench and before the strain, and on the
time spent under strain.
This type of behaviour is somewhat similar to what  could be expected in a simple Eyring description, with the yield stress being associated
with activated events of well defined energy which  could be a function of the waiting time \shortcite{RottlerRobbins2005}. However the coefficients
do not behave as expected in such a description, and in particular the coefficient $s_1$ tends to be constant rather than
proportional to temperature, which implies that somehow a different description of activation has to be found.

MD simulations have developed in several directions: attempts to get a better understanding of activation mechanisms
\shortcite{IlgBarrat2007,HaxtonLiu2007}, studies of strain localisation at intermediate scales, and  studies of correlation functions
that would allow the detection of microscopic dynamical heterogeneities.\shortcite{OnukiYamamoto1998,YamamotoKim2000}.
Studies of activation tend to support the notion of an effective temperature that would determine the rate at which the system
overcomes local energy barrier, and that depends on both temperature and shear rate.  Haxton and Liu \shortcite{HaxtonLiu2007}
claim that a data collapse for the flow curve can be achieved on this basis, with a single parameter $T_{eff} (T,\dot{\gamma})$.
They further argue  that this effective temperature is itself  given by the fluctuation dissipation
ratio in the system, determined  for various observables \shortcite{BerthierBarrat2002},\shortcite{OHernLangerLiuNagel2002}. A slightly different
approach was chosen in reference \shortcite{IlgBarrat2007}, in which an artificial "reaction coordinate" of a bistable system
was coupled to a system under shear, and an Arrhenius behaviour with a shear rate dependent effective temperature was reported. While this
points to the  existence of a mechanical noise that would complement--or even replace--the thermal noise, no progress has been made
that would allow one to relate the intensity of the noise to the shear rate. A clear justification of the SGR "effective temperature"
approach from microscopic simulations is therefore still missing.

 Persistent strain localisation in the form of "shear bands" has been observed in a relatively small number of MD simulation studies, in 2 as well as 3 dimensions.  In general such a persistent localisation is observed under strain conditions that are 
 not fully periodic (see however ref. \shortcite{ShiKatzLiFalk2007}, where Lees Edwards, fully periodic, boundary conditions are used) , but
 rather induced by boundaries (either in pure shear or uniaxial conditions). In \shortcite{VarnikBocquetBarratBerthier2003}, this localisation
 was observed at the walls of the simulation cell in isothermal simulations. Other examples of  strain localisation are obtained under  uniaxial
 loading with free boundaries \shortcite{ShiFalk2005,ShiFalk2006}, multiaxial loading \shortcite{BaileySchiotzJacobsen2004}, or nanoindentation \shortcite{ShiFalk2005a}
 sometimes  using a notch as initiator.
 These studies have not resulted in a clear understanding of the microscopic of shear bands,  as mentioned in\shortcite{BaileySchiotzJacobsen2004},
 "there is much that
is not understood about shear bands. This not only includes
why they form in the first place, but also what determines
their width, what distinguishes them structurally from the
surrounding material...". Simulation studies seem to indicate that some kind of initial heterogeneity
(e.g. boundaries) is important in the formation mechanism. Structural differences concerning the local bond order environment or local potential energy inside the shear band have been reported in some systems~\shortcite{ShiFalk2005,LiLi2006}, but they are not observed in all simulations~\shortcite{VarnikBocquetBarratBerthier2003}.
In the system studied by Shi and Falk~\shortcite{ShiFalk2005},
the quench procedure
was also shown to modify the ability to form shear bands. In rapidly quenched samples
	higher strain rates lead to increased localisation, while the more
	gradually quenched samples exhibit the opposite strain rate dependence.

Finally, we mention that indications of large scale dynamical  heterogeneities
have been found in early simulation work by Yamamoto and Onuki~\shortcite{YamamotoOnuki1998} on sheared systems of soft spheres.
These authors studied the spatial correlations between bond breakage events defined over a small, fixed time windows. Indications
of critical behaviour in the quasistatic limit are found in this work, which can
be seen as a precursor of more detailed studies of 4 point correlations. Such correlations  are studied in some more detail in
\shortcite{FurukawaTanaka2009} which shows for example anisotropy of dynamical heterogeneity in the non-Newtonian regime. There is
certainly, however, much room for a detailed study of dynamical heterogeneities in sheared systems at finite temperature as
a function of temperature and strain rate.

\subsection{The zero-temperature solid}

Molecular dynamics methods can thus mimic various features of material response which are commonly seen in experiments. But they are of course limited by the timescales they can access, which are orders of magnitude smaller than physical ones in numerical models of metallic glasses--the situation is not as dramatic if we seek to model e.g. colloidal glasses. Moreover, although simulations permit to access detailed information about the structure and dynamics, the microscopic motions are very blurred at finite temperature, so that it turns out to be quite difficult in practice to extract relevant information. 

This motivated interest for low-temperature systems: at finite but low temperature--lower than $T_g$--a glass would typically spend most of the time vibrating around a single local minimum in the potential energy landscape. One thus expects that this vibrational component make a trivial contribution while the most interesting aspects of dissipation in material response are controlled by hops between local minima--which may in some cases be thermally-activated. Therefore, following many studies  on glassy relaxation~\shortcite{Stillinger1995,DebenedettiStillinger2001,DoliwaHeuer2003,DoliwaHeuer2003a,DoliwaHeuer2003b}, it makes sense to try to separate vibrations from hopping in the PEL.

A lot of information about the mechanical response is thus captured by focussing on the $T=0$ limit. In the absence of an external drive, a material would just relax to a local minimum and rest. This limit, however, is very useful when studying the response to macroscopic deformation, as is helps focussing on deformation-induced changes of local minima~\shortcite{MalandroLacks1997,MalandroLacks1999}. 

\paragraph{Deformation-induced changes in the PEL}

Let us thus consider a numerical model of a glass at rest in a local minimum. This configuration is produced by energy minimization, starting from a configuration which may have had any thermo-mechanical history (slow annealing from a high-temperature state, or any form of plastic deformation).

We consider here the case when this minimum is stable and the applied deformation small enough to preserve stability. At zero temperature, the positions of the particles in the local minimum--that is as they adapt to the imposed deformation to preserve mechanical equilibrium--are smooth functions $\{\vec r_i(\gamma)\}$ of strain $\gamma$. In particular, these strain-induced changes are \emph{exactly} reversible; the system behaves as a perfectly elastic material.

Studies of elasticity in this regime date back to the works of Born and Huang~\shortcite{Huang1950,BornHuang1954}, who proposed to compute elastic constants by assuming that the relative displacements of all particles are affine, i.e. match the macroscopic strain. With this assumption, it is immediately possible to predict elastic moduli from the pair correlation function. The problem is that the particle trajectories $\{\vec r_i(\gamma)\}$, which trace strain-induced changes of a local minimum, are not simply dictated by macroscopic deformation. This approximation is only valid for very simple crystals, but not for glasses, and the displacement  $\{\vec r_i(\gamma)\}$ contains, in general, some non-affine contribution.

The existence of a non-affine component to the displacement field has been known for a long time, and its existence was hinted at in various papers as a trace of disorder~\shortcite{Alexander}. But it was really 
brought to the fore when it became clear \shortcite{LeonforteBoissiereTanguyWittmerBarrat2005} 
that it affects
 significantly the values of the macroscopic elastic moduli. The existence of a non affine field reflects the disorder in the local elastic moduli, which results in a heterogeneous elastic response exhibiting long ranged correlations \shortcite{DidonnaLubensky2005,Maloney2006}. This heterogeneity has been characterized in detail using a systematic coarse-graining approach~\shortcite{Tsamados2009}, with the results that below a scale of a few tens of particle sizes the
elastic constants (and especially the shear moduli) differ significantly from their macroscopic values.

Analytical expressions are available~\shortcite{MaloneyLemaitre2004a,LemaitreMaloney2006} for the non-affine field and for the corrected moduli. They correspond to the zero-temperature limit of those proposed by Lutsko at finite temperature~\shortcite{Lutsko1988}. The important point is that the $T=0$ non-affine field takes  the form:
\begin{equation}
\label{eqn:nonaff}
\frac{d \vec r_i}{d\gamma}= -\mathcal{H}_{ij}^{-1}.\vec{\Xi}_j
\end{equation}
where ${\mathcal H}$ is the Hessian matrix, and $\vec{\Xi}$ a vector field corresponding to infinitesimal, strain-induced, changes of forces on each particle. As $\vec{\Xi}$ can be constructed from the derivatives of the potential function, it is easy to show that it does not vanish or present any singular behaviour. Consequently the non-affine displacements will 
acquire any singular contribution only from the
inversion of the
of a Hessian matrix. Since glasses present many low-lying modes, the non-affine field will pick up information about the existence of soft regions in space~\shortcite{Papakonstantopoulos2008,Mayr2009,Tsamados2009}.
 These soft regions are precisely those that
 control plasticity, when a material is driven by strain towards instabilities~\shortcite{LemaitreCaroli2007}.

Although it is, in principle, outside the scope of this review, we briefly mention the special case of granular systems
interacting through repulsive contact forces. At zero temperature, these systems lose their rigidity abruptly as the density is decreased,
which defines the (un)jamming transition. In the vicinity of  the jamming density, the  non affine, heterogeneous  response becomes the  dominant feature in the elastic deformation of such systems \shortcite{VanHecke2009}, and can be associated with a diverging "isostatic" length scale, below which the stability is governed by boundary conditions \shortcite{Wyart2005}.

\paragraph{Plastic events in AQS shear}

The AQS (athermal quasi-static) protocol consists in applying quasi-static deformation to the zero-temperature solid described previously. It is implemented by a two-step protocol~\shortcite{MalandroLacks1997,MalandroLacks1999,MaloneyLemaitre2004b,MaloneyLemaitre2006}. Starting from an equilibrium configuration: (i) the system is deformed homogeneously by a very small increment; (ii) energy is minimized. If the increments are sufficiently small, the system will be able to track continuously the deformation-induced changes of the occupied minimum. As the procedure is iterated, the local minimum will eventually become unstable. Minimization will then let the system relax to a new configuration which is disconnected from the previous one. On a flow curve, as illustrated on Fig.~\ref{fig:stress-strain}, the AQS response shows up as a series of continuous branches--corresponding to the reversible tracking of single minima--and discontinuous jumps--corresponding to ``plastic events''.

 Given that the elastic response is perfectly
reversible, the plastic events account exactly for all the dissipation,
and they can be identified
unambiguously from the discontinuities of the stress curve. This has made possible  detailed studies of 
their organization in space and of their size distribution~\shortcite{MaloneyLemaitre2004b,TanguyLeonforteBarrat2006,MaloneyLemaitre2006,BaileySchiotzLemaitreJacobsen2007,LernerProcaccia2009}.

In some cases, single shear transformations can be observed. This can be done by looking at events of small sizes, which are present in the steady state flow~\shortcite{MaloneyLemaitre2004b,MaloneyLemaitre2006}, but more easily found during the early loading phase from an annealed, isotropic state~\shortcite{TanguyLeonforteBarrat2006}. Another way to access them is to look at the onset of plastic events, which were found to involve a single eigenvalue going to zero~\shortcite{MalandroLacks1997,MalandroLacks1999,DoyeWales2002}. Close to instability, the non-affine field (see Eq.~(\ref{eqn:nonaff})) aligns with the vanishing mode and the singular behaviour of energy, stresses and moduli can even be predicted~\shortcite{MaloneyLemaitre2004a}. This has allowed one to study the spatial structure of this mode~\shortcite{MaloneyLemaitre2006,TanguyLeonforteBarrat2006}, showing that it generically presents the quadrupolar structure and a decay away from its center,both of  which are predicted by the Eshelby inclusion model~\shortcite{Eshelby1957,PicardAjdariLequeuxBocquet2004}.

But in steady flow--that is past some preparation-dependent initial transient~\shortcite{TanguyLeonforteBarrat2006}--plastic events are not in general composed of single zone flips but typically involve many of them collectively organized as avalanches~\shortcite{MaloneyLemaitre2004b,MaloneyLemaitre2006,BaileySchiotzLemaitreJacobsen2007,LernerProcaccia2009}. This last claim is supported, in particular, by measurements of the average size of stress (resp. energy) drops, which scale as $L^\beta$ (resp. $L^\alpha$) with system size $L$, and $\alpha,\beta<1$. Despite some differences in the reported exponents, this power law scaling is now accepted as a fact, and proves that the local rearrangements are strongly correlated, in notable contradiction with the mean-field assumption.

\paragraph{Avalanches at finite strain rates}

An immediate question is whether avalanches are only a feature of the quasi-static limit, or whether they exist for realistic values of the external parameters. In particular, the unfolding of each avalanche should take some time, determined by the duration of elementary flips and by the propagation times of elastic signals   between flip events~\shortcite{LemaitreCaroli2009}. Even if we stick with athermal systems, as soon as a finite strain rate is introduced, the avalanches may start to overlap because of their finite duration. Hence they can no longer be properly identified as in the quasi-static limit.

So, a first problem when going away from the AQS limit is to define valid observables which allow one to characterize the existence of an underlying avalanche process, i.e. of correlations between local rearrangement events. A protocol has thus been designed to characterize avalanches in 2D athermal systems, from measurements of transverse diffusion~\shortcite{LemaitreCaroli2009}. The principle starts from the observations that in AQS simulations, the transverse diffusion constant exhibits strong size dependence~\shortcite{LemaitreCaroli2007,MaloneyRobbins2008}. This dependence can then be attributed to the organization of Eshelby flips along roughly linear patterns~\shortcite{LemaitreCaroli2009}, so that the diffusion constant can be expressed as a function of a typical avalanche size.

This observations has led to the proposal that the avalanche size should depend on the strain-rate as $\ell\propto1/\dot\gamma^{1/d}$ in dimension $d$~\shortcite{LemaitreCaroli2009}. As $\dot\gamma$ decreases, the avalanche size should saturate at a length-scale $\propto L$ below some critical strain-rate $\gamma_c\propto1/L^{d}$, as in any usual cross-over. This suggestion is based on an interpretation of the avalanche size as being limited by the screening of Eshelby elastic signals by the background noise due resulting from all the flips in the system.

These scalings are consistent with the rough scaling of the average stress drop as $\langle \sigma\rangle\propto 1/L$ found in~\shortcite{MaloneyLemaitre2004b,MaloneyLemaitre2006}, but of course, these estimates remain rough, and it is not ruled out that more precise measurements would provide slightly different exponents consistent with the values of $\beta$ found in~\shortcite{LernerProcaccia2009}. What should remain, however, is that avalanches are present at all physically accessible strain rates, even when they overlap in time,
and even when they cannot be accessed
 via the identification of separated events.

\paragraph{Steps towards a phenomenology of plasticity}

The observation of avalanches, and their properties in the quasi-static limit and at finite strain rates provides clear benchmarks for future theories of plasticity. Yet, as usual in studies of amorphous systems, we can observe and characterize the avalanche process; we can conclude that some form of correlation exist between flip events; but the underlying mechanisms which promote these correlations remain quite difficult to identify. In fact, several processes must  occur together to make the avalanche behaviour possible.

First, the flip-flip interaction must be mediated in some way. Here, the medium is elastic and this is know to produce long-ranged effects. The Eshelby mechanism~\shortcite{Eshelby1957,Argon1979,BulatovArgon1994a,BulatovArgon1994b,BulatovArgon1994c} must play a central role, namely, each flip alters the stress in its surroundings, which may push a nearby region past its instability threshold, hence trigger a secondary instability. The existence of this mechanism is supported by the observation of single flips in AQS simulations~\shortcite{MaloneyLemaitre2004a,MaloneyLemaitre2004b,TanguyLeonforteBarrat2006}, by the measurement of the stress decay in space~\shortcite{MaloneyLemaitre2006}, and by direct visualisation of flips events at finite strain rates~\shortcite{LemaitreCaroli2009}. There are also now direct observations that a primary zone flip can push a nearby one closer to instability~\shortcite{LemaitreCaroli2007,LemaitreCaroli2009}, thus showing that the Eshelby mechanism is fully at work.

But we must note also that when a system is sheared from carefully annealed, isotropic, state zone flips tend to be isolated instead of organizing as avalanches~\shortcite{TanguyLeonforteBarrat2006}. There is no reason why there should be any fundamental difference between the basic mechanisms which are at work during the loading phase or in steady state. Therefore, the difference between the early stage response and the steady-state flow must indicate that the state of the material evolves under loading, in a way which increases the density of near-threshold, soft, zones, consistent with the idea that strain results in progressive advection of the  zones towards their instability thresholds~\shortcite{LemaitreCaroli2007}. In steady state, the density of near-threshold regions is high enough, and the Eshelby stress redistribution operates efficiently. In early loading, the density of near-threshold regions would be lower if the system is carefully annealed, so that isolated events can be more easily identified.

Like in a game of dominoes, the Eshelby mechanism makes the avalanche process possible. But it can occur only if the density of near-threshold regions is high enough, which must be a property of the material structure. The question thus 
 re-emerges of how to characterize
 the regions or ``zones'' where elementary shear transformation may occur. Could we identify them a priori? What would be their density? Do they correlate with some property of the local structure--stress, density, moduli?  Up to now, the particle based studies that have addressed these questions have resulted in rather disappointing results. First of all, it appears that the regions in which the localized plastic events take place are not, in general, under particularly high stress. More precisely, the probability of observing a yield event in a region under high stress is indeed higher, but this is balanced by the fact that
the number of such regions is small. A good correlation, on the other hand, has been established between yield events and regions with low values of elastic moduli \shortcite{Mayr2009,Tsamados2009}. This suggests that the heterogeneity of local elastic constants should be taken into account in more coarse-grained models. Unfortunately, the local elasticity is already a rather complex property, and attempts to directly relate the probability of yielding to the local atomic structure have not been very successful, although a correlation
with the shape of the Voronoi volume was observed in polymer glasses \shortcite{Papakonstantopoulos2008}. In systems with strongly directional bonding such as amorphous silicon, a correlation could also be established between the density of bonding defects and the local plastic activity \shortcite{TalatiAlbaretTanguy2009}. One must acknowledge that a link is still missing that would allow a better control of plastic properties directly from the design of the microscopic structure.

\section{Perspectives}
\label{sec:3}
We close this review by emphasizing a few key issues which, in our opinion,  have to be addressed in order to build a consistent theory
of amorphous solids under strain, including the heterogeneous, fluctuating aspects. Most of these points have been discussed in detail in the previous sections.

The present consensus on the existence and importance of "zones" and "flips" as the essential building blocks
of the plastic activity makes it strongly desirable to have a better understanding of these zones from the standpoint of the
local microstructure. Many more atomistic simulations, involving efficient sampling techniques allowing for longer simulation runs \shortcite{RodneySchuh2009b},  and
using various types of interatomic potentials,  will be needed to achieve such an understanding. It might even be the case that a predictive  search for
microstructural characteristics of flipping zones is illusory, and that the local plastic activity is a result from so many factors that it is essentially unpredictable.

Even if the flips are not associated with well defined zones at the structural level, the essential features of the current models of elasto-plastic behaviour
are in fact statistical in nature. It is therefore essential to develop tools that allow one  to quantify in an unbiased, statistical manner the plastic
activity, so that a comparison between models, numerical simulations and experiments is possible. Such a strategy has proven very successful in the field of glassy systems and supercooled liquids at rest, and should be extended to the case of low temperature, driven amorphous systems. In particular, a statistical description of dynamical heterogeneities in strained systems, a quantification  of avalanche distributions (in energy and size), and of the relevant correlation lengths, is still missing. The influence of temperature and strain rate on these quantities should also be a subject of interest.

In the introduction, we insisted on the similarities between "soft" systems probed by rheological experiments, and "hard" systems" such as metallic glasses. The similarities are useful and important in terms of theoretical modelling, still the practical applications are very different. In soft matter, the focus will be on steady state or low frequency rheological behaviour, for which a permanent avalanching regime can be established, and the memory of the initial state is wiped out after a few cycles. In contrast, hard materials undergo irreversible failure after a few  percent of strain, so that a permanent regime cannot be established, and the thermomechanical history of the initial state becomes of crucial importance. Adding the system history as an additional "variable" extends considerably the complexity of the problem, so that with the  exceptions of a few studies \shortcite{UtzDebenedettiStillinger2000,ShiFalk2005,RottlerRobbins2005} most simulation works have focused on steady state properties. 
Therefore the influence of thermomechanical history, and the relevance of the correlated avalanches mechanisms 
in term of material failure, remain outstanding questions in which simulations studies could be compared to experimental results and theoretical approaches \shortcite{FalkLangerPechenik2004}.

We also mentioned that ``hard'' and ``soft'' glasses may differ broadly in terms of the relevant "reduced" parameters, that is after mass, length, and energy scales are made dimensionless. Colloidal glasses, for example, involve low reduced temperatures and high reduced strain-rates while metallic glasses involve converse conditions. If recent numerical works have focused on athermal systems, it is because the absence of thermal fluctuations facilitates the observation of elementary mechanisms of deformation. Indeed, it takes only quite small temperatures--compared to $T_g$--to induce (high-frequency) fluctuations which are significantly larger than the changes associated with plastic deformation and energy dissipation~\shortcite{HentschelKarmakarLernerProcaccia2010}: this of course limits our capacities of investigation. Understanding whether and how the mechanisms identified at zero temperature carry over to finite-$T$ systems will inevitably become one of the major theme in this field. As often in the 
field of amorphous systems, rather different approaches can be developed, depending on whether one starts  starting a ''low temperature''  or from a ''high temperature'' viewpoint. Examples could
be elasto-plastic models on one hand, and the mode-coupling theory (which we mentioned only briefly) on the other. Trying to understand the range of applicability of such different approaches, and possibly to get a consistent (if not unified)  view of their relevance for various   experimental systems and conditions remains a real challenge.

Finally, we made in the introduction a distinction between the heterogeneities associated with statistical fluctuations of a globally homogeneous strain,
and the "macroscopic", long lived heterogeneities described as strain localisation. The description of the latter situation has made some progress recently, with the realization that a  mechanism involving the diffusion of some auxiliary state variable (fluidity, effective temperature, free volume) was
in general needed to produce such heterogeneities. This auxiliary variable could even be related to some of the statistical aspects mentioned above. However, all these approaches are still oversimplified  (scalar stress variable, absence of convection...) so that  direct comparison with  a realistic experimental geometry is difficult.

{\bf Acknowledgements: } We acknowledge many useful discussions and collaborations with 
Lyd\'eric Bocquet, Christiane Caroli, Peter Sollich, Anne Tanguy, Michel Tsamados,
and critical comments on the manuscript by Micahel Falk and Jim Langer.

\bibliographystyle{OUPnamed_notitle}
\bibliography{biblio-final}
%


\end{document}